EL 3257
**Games as a Platform for Student Participation in Authentic Scientific Research**

**Rikke Magnussen[1], Sidse Damgaard Hansen[2], Tilo Planke[2], Jacob Friis Sherson[2], AU Ideas Center for Community Driven Research, CODER**
1, ResearchLab: ICT and Design for Learning, Department of Communication, Aalborg University, Denmark
2, Department of Physics and Astronomy, Aarhus University, Denmark
rikkem@hum.aau.dk
sdh06@phys.au.dk
tilo@phys.au.dk
sherson@phys.au.dk

**Abstract**: This paper presents results from the design and testing of an educational version of Quantum Moves, a Scientific Discovery Game that allows players to help solve authentic scientific challenges in the effort to develop a quantum computer. The primary aim of developing a game-based platform for student-research collaboration is to investigate if and how this type of game concept can strengthen authentic experimental practice and the creation of new knowledge in science education. Researchers and game developers tested the game in three separate high school classes (Class 1, 2, and 3). The tests were documented using video observations of students playing the game, qualitative interviews, and qualitative and quantitative questionnaires. The focus of the tests has been to study players' motivation and their experience of learning through participation in authentic scientific inquiry.
In questionnaires conducted in the two first test classes students found that the aspects of doing "real scientific research" and solving physics problems were the more interesting aspects of playing the game. However, designing a game that facilitates professional research collaboration while simultaneously introducing quantum physics to high school students proved to be a challenge. A collaborative learning design was implemented in Class 3, where students were given expert roles such as experimental and theoretical physicists. This significantly improved the students' feeling of learning physics compared to Class 1 and 2.
Overall the results presented in this paper indicate that the possibility of participating in authentic scientific experiments, which this class of games opens, is highly motivating for students. The findings also show that the learning design in the class setting must be considered in order to improve the students' experience of learning and that various design challenges remain to be addressed even further.

1. **Introduction**
The use of games and simulations in science education was introduced in the 1970s. In the early 1980s the potential of games and simulations as new teaching tools was extensively discussed (Ellington et al., 1981). In the early 1990s the first IT-based educational games for exploration of the natural sciences and technical subjects were developed (Egenfeldt-Nielsen, 2005). Since the beginning of the new millennium, there has been an increasing awareness of the opportunities that new types of commercial computer games can offer science teaching, and of developing new formats framing aspects of real-life science learning environments that allow players to tackle complex problems in simulated professional contexts (e.g. Squire &

Klopfer, 2007; Magnussen, 2007). These games supported new practices in science education, such as student development of new professional inquiry tools for imaginative creation and representation of new knowledge. These new practices are influenced by non-school factors such as fictional characters and authentic professional tools (Magnussen, 2008). More recently, scientific discovery games have been developed in which player contributions to real-life research challenges are an integral part of the game. In this class of games complex scientific problems are translated into puzzles and a game-like mechanism is provided for non-expert players to help solve the presented problems (Cooper et al., 2010).

The current paper presents the design considerations and test results of the adaptation of a new scientific discovery game for in-school learning environments. Some of the main design considerations include making the complicated scientific background understandable, getting the students engaged in the process of generating new knowledge, and creating a sufficiently strong connection from the research challenge to the core student curriculum.

## 2. Background: Gamified research collaboration in science education

One of the main focuses in the development of science game formats over the past 10 year has been how the medium of games can introduce new approaches to authentic science education (Gee, 2003; Magnussen, 2008). Concrete examples are the so called Profession Simulation Games or Epistemic Games. Prime examples of these are profession simulation games that simulate objectives and environments of a specific profession by using the technology, tools and/or methods of the profession.

### 2.1: Profession Simulation Games

The objective of Profession Simulation Games or Epistemic Games is to draw on the game medium as a framework for the complex interdisciplinary creation of new knowledge and innovative solutions. Examples of this type of games are students playing engineers identifying polluted sites (Squire & Klopfer, 2007), urban planners planning a central urban area (Shaffer, 2006), or forensic experts investigating murder cases (Magnussen, 2007). Profession Simulation Games are designed to meet the official educational objectives (Magnussen, 2008), while the aim of Epistemic Games are:

> "…to create the epistemic frame of a socially valued community by re-creating the process by which individuals develop the skills, knowledge, identities, values and epistemology of that community" (Shaffer, 2007, p. 164),

The motivation for developing this type of games stems from a critique of the teaching of standardised skills to children in today's school system (Gee, 2003). It is argued that few schools teach students how to create knowledge; instead, students are taught that knowledge is static and complete. This means that they become experts at consuming rather than producing knowledge (Sawyer, 2006). As a result, the medium of games has been used to create environments with simulations of complex real-life situations, where students have to think like professionals and solve problems in innovative ways within the context of the knowledge domain of the profession defined by James Paul Gee as:

> ''Any actual domain of knowledge, academic or not, is first and foremost a set of activities (special ways of acting and interacting so as to produce and use knowledge) and experiences (special ways of seeing, valuing, and being in the world). Physicists do

> physics. They talk physics. And when they are being physicists, they see and value the
> world in a different way than do non-physicists." (Gee, 2005, p.1)

Interesting issues arise, however, in relation to this class of games that need to be addressed when discussing the integration of creation of new knowledge and authentic science practice in science education. Even though the games integrate professional values and tools, they remain *simulations* of professional practices. This aspect of the games brings up the matter of whether students learn to work *as* a scientific expert or whether they learn how to *be* a scientific expert. This may depend on various design elements of Profession Simulation Games. First, the clients and experts that students collaborate with in the games are fictional characters with fictional problems that do not necessarily have relevance outside school. Secondly, the fictional problems to be solved in these games often follow a linear path and have a clear starting and ending point. This is clearly different than real-life professional problem solving, where the processes are more multidimensional. Finally, even though these types of games have been shown to support student creation of new process tools, the solutions are often pre-defined and already known by the teachers (Magnussen, 2008). This stands in contrast to the real-life open-ended tasks professionals face which can be carried out in various ways. Here, the chance of success or failure is always an issue of consideration. This raises the question: whether students, in spite of the games involving authentic values and tools from a given professional practice only learn about scientific expertise in a scenario adapted to the school context or whether they experience being part of a knowledge domain as participants in a scientific process that includes an authentic professional context.

## 2.2 Scientific Discovery Games

The development of the so-called Scientific Discovery Games within the past couple of years brings new elements into the issue of game-based participation in a knowledge domain. The main goal of this class of games is to create a platform that enables and motivates players to contribute in solving scientific problems (Cooper et al., 2011). The most well recognised example of this class of games is Foldit which is an online puzzle game where players participate in folding amino acid chains to form new protein structures. Presented with a primary protein sequence or partially folded structure, players need to find the – often unknown - lowest-energy three-dimensional structure. Players manipulate the protein structure by pulling, twisting and tugging the protein backbone and side chains into various configurations (Good & Su, 2011).

Scientific discovery games contain specific design features that distinguish them from the majority of other games (Cooper et al., 2010; Good & Su, 2011). First, Scientific Discovery Games are designed for non-expert players to advance a scientific domain. As a result, the visual features and graphics must enable beginners to experiment with highly complex solutions and scientific information. This requires that the game interface is designed to gradually introduce beginners to a highly complex field while simultaneously motivating them to play the game. Another distinctive feature of this class of games is that the problems do not have pre-defined solutions; even the game designers do not know the answers, of which there are potentially more than one. This also implies that the interactive design must make exploration and experimentation processes possible while simultaneously respecting real scientific constraints. Consequently, the scoring mechanism must reward several distinct player strategies while remaining true to the latest knowledge about the scientific phenomenon (Cooper et al., 2010; Good & Su, 2011).

The goal of the research effort presented in this article is to expand the Scientific Discovery Game concept to be included in an educational context. An educational design has been created where students gain first-hand experience with creating new scientific knowledge in collaboration with researchers. The design aims to create the experience that this knowledge is not static, but constantly evolving. However, there are educational challenges in designing this type of games. Scientific Discovery Games have the potential to introduce authentic experiments and creation of new scientific knowledge into science education. Therefore, it is crucial to understand if and how the game environment contributes to students' experience of participating in research. Moreover, we also need to understand the main motivating factor for playing this type of games in school; is it competing against other students, contributing to science, or collaborating with scientists? Understanding the perceived learning of students and implications of introducing this type of games to different types of students is thus another focus point in the research presented in this paper. A final challenge for the educational designers is the observation that student perception of learning is strongly affected by the success of relating the context of the game to the core curriculum. In this paper we present the design and initial test results of the adaption of a Scientific Discovery Game, Quantum Moves, to a school environment in order to teach quantum physics to high school students in Denmark. We furthermore discuss the potential and challenges of designing this type of games for science education.

### 3. The Game Quantum Moves: Gamified Quantum Computer Research

The game, Quantum Moves, represents a collaborative effort between researchers in physics, computer science and game-based learning at the interdisciplinary AU Ideas Pilot Center for Community-Driven Research, established January 2012. The focus of the quantum game project is the research-based production of a game-based platform for player participation in quantum research, more specifically in the research of building a quantum computer. In the following the concept of quantum computers and the mission of Quantum Moves are presented briefly. Although it is not the intention of this paper to give a detailed account of the online game, some physical details are included here to give an insight into the new concepts that students were exposed to during the interventions into the high school classes. The detailed considerations concerning the optimal didactic framing of the intervention are discussed in chapter 4.

### 3.1 Quantum computers and the AU candidate

All data in regular computers are stored in vast amounts of small electrical devices called transistors, each representing either the value 0 or the value 1. In a quantum computer (QC) these so-called bits are replaced by individual quantum systems such as atoms, electrons, ions, or photons. Due to the fundamental principle of superposition these *quantum bits* can be both 0 and 1 simultaneously. This means that many calculations can be performed in parallel and even a relatively small quantum computer would contain more computational power than all conventional computers combined. Very small QCs have existed for over a decade but to date the world record is a mere 14 quantum bits.

The building blocks in the Aarhus University (AU) candidate for a large quantum computer are extremely cold atoms arranged in a periodic pattern similar to eggs in an egg-tray. The physicists have recently participated in the experimental creation of such a crystal containing hundreds of atoms (Sherson et al., 2010) and the remaining challenge is to figure out how to perform operations on this system. Recent theoretical work has indicated this to be possible by using a tightly focussed laser beam (a tweezer of light) to pick up individual atoms and transport them around in the crystal (Weitenberg et al., 2011). This movement, however, poses a large challenge because any sudden movement will cause the atoms to slosh in the tweezer. This sloshing will translate into errors in the calculations and the basic unsolved challenge is therefore to find the best way of moving the tweezer containing the atom to minimize the final sloshing.

**3.2 The Scientific Discovery Game Quantum Moves**
The researchers have tried out multitudes of different motional paths in extensive computer simulations but believe that they have not yet found the best solutions for moving the atoms in the AU candidate for a QC yet. To solve this problem, they developed the Scientific Discovery Game, Quantum Moves, in which players try out their solutions in a graphical simulation interface. A particular path consists of a choice for the position and intensity of the tweezers. For each attempt a player makes, a score is calculated based on the quality of the resulting quantum computer. The performance of every player is logged centrally and the overall highest score will always correspond to the state of the art of the research field and can thus be improved as people play. We hope that the approach will be effective not only due to the sheer quantity of potential players but also because players can potentially apply the distinctly human skills of pattern recognition and intuition to perform a much more intelligent optimisation than computers can.

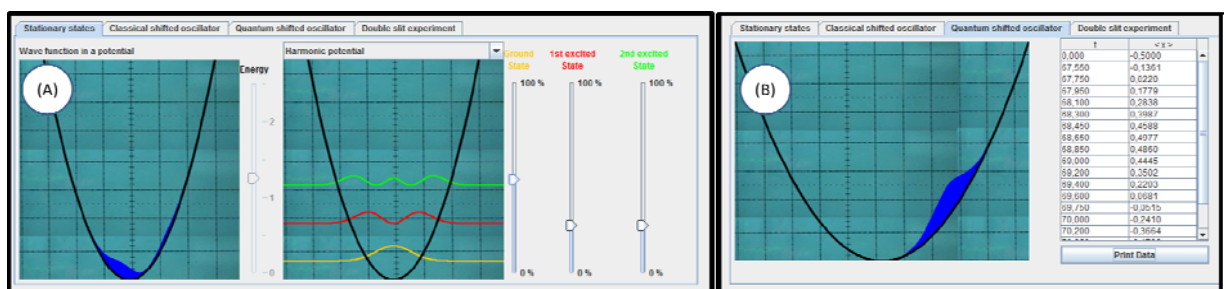

**Figure 1:** Examples of tutorial games to introduce the quantum mechanical concepts and methods needed to understand the scientific challenge. (A) An illustration of the allowed quantum mechanical states and sliders to create mixtures of these. (B) A small tutorial game teaching the student how to remove kinetic energy from an oscillating atom.

The educational version of Quantum Moves (EQM) has two parts; a tutorial part and a scientific part. The tutorials consists of games and simulations that introduce players to quantum physics (figure 1) and where the game mechanics of moving the atom is introduced (figure 2 A). Examples of actual scientific research games in different iterations are shown in figure 2 B and figure 3. Figure 1 (A) illustrates the allowed quantum mechanical states and offers sliders to create mixtures of these. The basic lesson is that if the atom (represented as a liquid-like substance inside the graph) is purely in one of the allowed states of the well, it will not move in time but in a mixture it will. In figure 1 (B) an atom is displaced in the well and therefore starts oscillating. The user then has to remove the kinetic energy by moving the well from side to side. The data on the position of the atom versus time is listed to the right, allowing students to transfer the information to a plotting program to analyse the results of their experiment.

**3.3 The iterative adaption of Quantum Moves to high school interventions**
The first types of tutorials – as the ones show in figure 1 – are the most important in this context because they were developed and iteratively modified with the specific purpose of establishing as close a link between the physics of the research question and the core curriculum in high school interventions. At the end of each intervention the students were then exposed to the version of the online game at the time of that particular intervention. This explains why the students were exposed to two different versions of the online game (figure 2 B and figure 3). One example of the iterative design is the addition of the possibility of extracting the data of the position of the atom versus time as shown on the right hand side of Figure 2 B. In initial interventions the position of the atom at various times was simply visualised in a graph to the right of the game window. Later we added access to the actual data because we found that the process of extracting the data, importing it into a well-known program, and plotting it there created a much more active learning atmosphere in the class room.

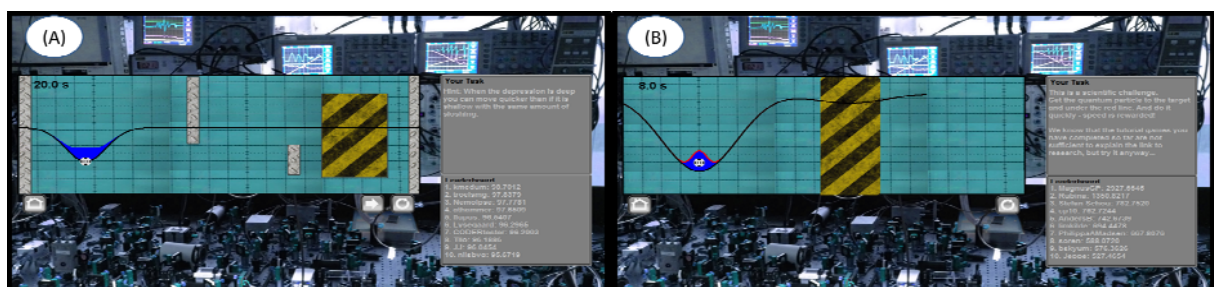

**Figure 2:** Screen shots from the online Quantum Moves game (A) An introductory challenge where players must transport the atom to the striped target area without hitting the walls. (B) The real scientific challenge where players must deliver the atom in the target area without any sloshing.

**4. Investigating the game in high schools**

The EQM platform has been tested in Danish high schools with students aged 17 – 20. The focus of these tests has been to investigate the following research perspectives. First of all, the aim of the study is to understand which new practices this class of games introduces into science education. Here, we focus particularly on investigating if and how it strengthens students' perception of participation in authentic scientific research. Secondly, we aim to understand students' motivation for playing scientific research games and whether the motivation is intrinsic (related to game factors) or if students are extrinsically motivated by factors for instance related to the research participation. Thirdly, the studies have focused on investigating the students' perceived learning: What influences their experience of learning physics from the game and how does the game initiate new learning approaches for different student groups. These research perspectives have been tested in cycles of research and development which will be described below.

**4.1 Methodology and test design**

The methodology used in the development of the didactic components of EQM followed a design-based research process and involved various design cycles, interventions, analyses and redesign (Brown, 1992). In design-based research the focus is both on developing learning environments as well as domain specific theories (Cobb et al., 2003). The method serves as a methodological framework for integrating differing methods at the various stages of research and development (Squire, 2005). In the first phase of design of and intervention with the game, focus was on understanding whether it was intrinsic game factors, e.g. elements of competition, that motivated the students to play the game or extrinsic factors, e.g. research participation, that was the driving force. A beta version of the game (see figure 3) was completed in early 2012 and interventions were conducted in two high school classes. Class 1 consisted of 20 students 17-19 years of age in their second year of high school, while Class 2 comprised 20 students 17- 20 years of age in their third and final year of high school. In both classes the teacher introduced basic quantum physics principles (as those discussed in Section 3 and Figure 1) in the lectures preceding the intervention. Quantum researchers from the development team introduced EQM to the classes along with their view of the link between the curriculum and the research challenge. Based on the subsequent discussions it was a clear observation that the researchers could present the particular parts of the curriculum much more engagingly and convincingly than the teacher. We speculate that the reason could be the perception that the researcher represents a living field of knowledge with a close connection to the curriculum and the research frontier, whereas the teacher represents a static and artificial collection of facts. Following the introduction, the test of the different sub-games was conducted over two class periods. Students' game participation was documented using video observations combined with short semi-structured qualitative interviews (Kvale, 1996) of students and teachers during play sessions. The aim was to identify new types of practices in the interaction with the game. In addition, qualitative interviews were conducted with the entire class and their teacher after each play session. The classes were also asked to fill out a written questionnaire after the test, answering qualitative questions such as "What was the best/worst part of playing the quantum game?" and "How does playing the quantum game differ from your other physics teaching?". Quantitative questions included "Rate the following - Have you learned physics by playing the game?" where students were asked to rate the statement on a scale from 1 (not at all) – 5 (a lot).

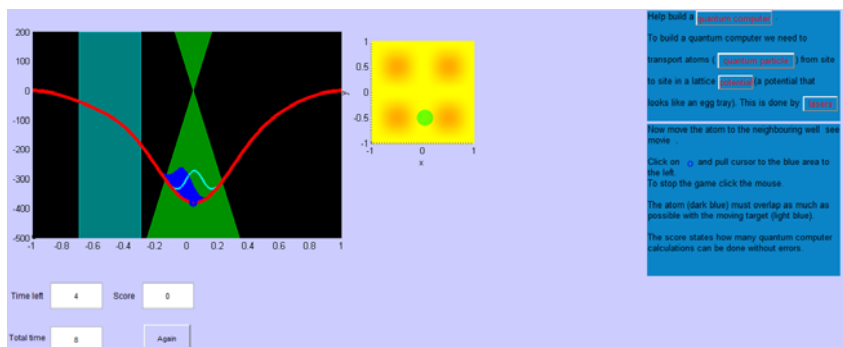

**Figure 3:** An early alpha version of Quantum Moves tested in Class 1 and 2. Players have to move the atom from a well on the right to a well on the left while agitating the atom as little as possible.

Focus of these two first tests was to collect data on motivational aspects of playing the game and the students' experiences of learning quantum physics from the game. In the second design-based research cycle of problem definition, design, intervention and analysis focus was on strengthening the students' experiences of learning quantum physics. Based on results from the initial tests, the game was further developed and a new version was tested in April 2013 in a high school in west Denmark (Class 3).
With the aim to strengthen the student's experience of participating in and learning authentic professional quantum physics practice a didactic design was developed in collaboration with the class teacher. The focus of the design was to combine theoretical approaches, experimental practice using physical tools, and play with game simulations. The goal was to boost the authentic aspects as well as the theoretical and experimental practices by assigning students roles in the game as experts. This design was inspired by elements of simulation games on specific professions (Magnussen, 2007) and notions of collaborative learning (Dillenbourg, 1999). The game test session in Class 3 was conducted over four school periods and had been introduced in previous lessons by the teacher. At the beginning of the first lecture the students were introduced to a simulated professional setup where they played the roles of a research team of physicists working on the development of a quantum computer by optimising the laser movement to transport atoms. Students were divided into three different teams, each one representing an area of expertise. One group comprised experimental physicists who were working to better understand the transport movement by doing experiments on a classical analogue to the quantum problem. The second group of experts were IT professionals specialised in simulations, who performed virtual experiments in the games. The last team worked theoretically and focused on understanding the mathematics behind the various elements of the atomic transport. The teacher assigned each student to the particular group best matching their interest and skill level. Students worked in their group for three class periods. They were then mixed with students from the two other groups to share observations and to produce a poster or film with their conclusions. The sessions were documented through video observations, questionnaires and qualitative interviews similar to the studies in Class 1 and 2.

**4.2 Results Tests Class 1 and 2: Motivation and Authenticity**
Similar observations were made in the two high school classes, Class 1 and 2, where the game was first tested. Students in the two classes were initially intensely interested and motivated and worked continuously with the game during the two class periods. At the end of the second class period, some students had begun to lose interest. This was especially the case for the youngest group of students from Class 1. The test was set up so that the students had to download the game on their own laptops prior to the test session. The majority of the students used their own laptops, but a few shared laptops in groups of two or three. In both classes, students discussed strategies and patterns for how to transport atoms in different environments while playing. They also discussed their scores and commented on results, high scores eliciting cheering. Scores seemed to be a central motivating factor for the students. However, results from the qualitative interviews after the test and written questionnaire showed that other elements contributed as motivating factors as well. The answers from the interviews and the questionnaire varied according to what participants thought was the best part about playing the game.

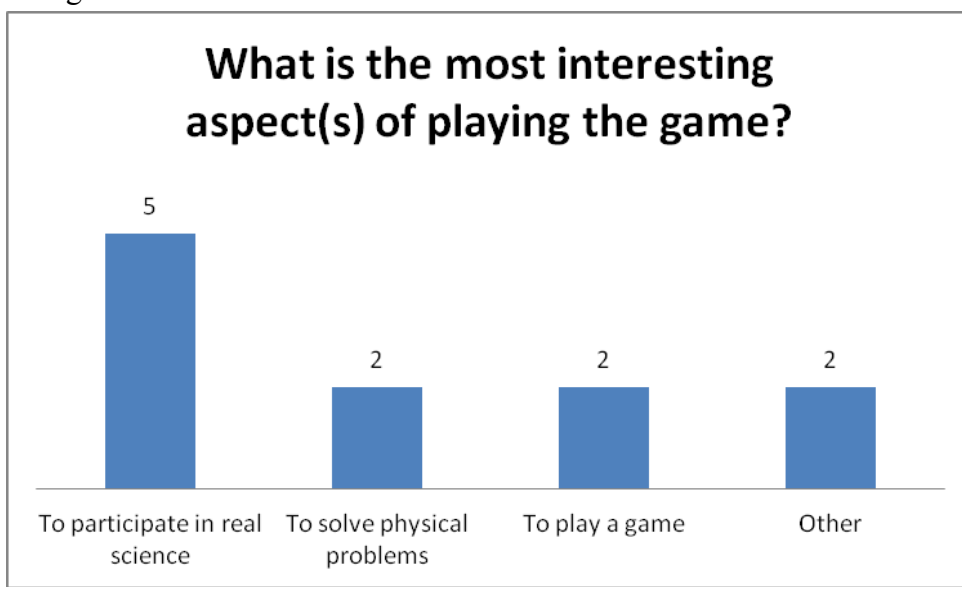

**Figure 4:** The answers in Class 1 on the question "What is the most interesting aspects(s) of playing the game". Seven students answered the questionnaire. The students were allowed to choose more than one of the four options.

The questionnaire was unfortunately completed by only 7 out of 20 students in Class 1. The results therefore carry no statistical significance in themselves but are included to exemplify student opinions. On the question "What is the most interesting aspect(s) of playing the game?" 5 of the 7 of students in Class 1 answered "To participate in real scientific research", the other options where each chosen by 2 of the students, see figure 4. These results show that research participation was the primary motivational factor in this class.

In the qualitative interview with the entire class, the students were asked what the biggest difference between playing the game and normal teaching was. One student stated that, "In the normal teaching you only calculate the results, while in the game you get the feeling of directly doing the experiment". When students and (after the interviews with the students) teachers were asked to expand on this comment, they explained that what happened in the game felt like an experiment compared to class lab work, which they saw as mere demonstrations. This indicates that students exposed to EQM experience participating in actual quantum physics experiments in sharp contrast to the highly theoretical level at which this subject is conventionally taught in Danish high schools.

Class 2's answers to questions about motivational factors differed slightly from the answers in Class 1 but overall showed similar results, see figure 5.

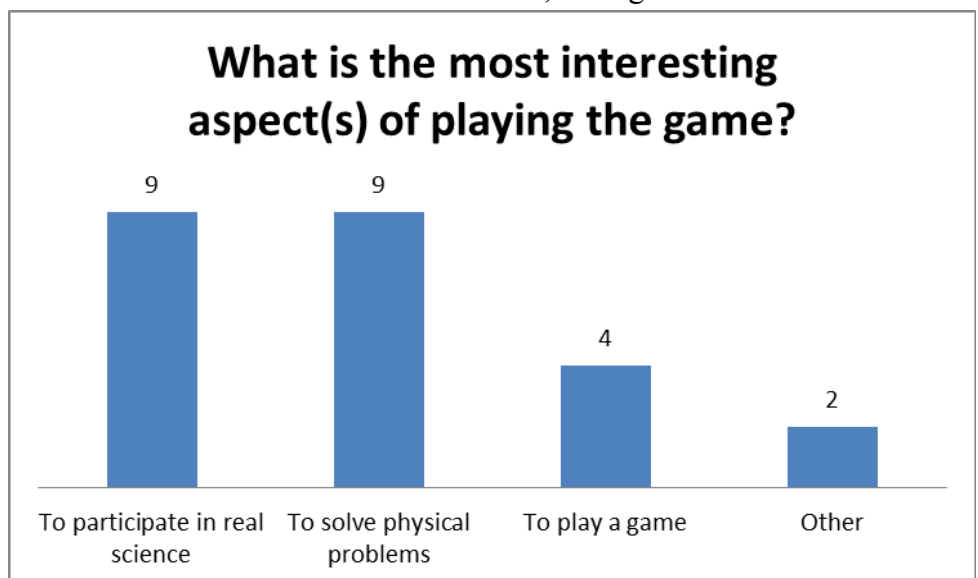

**Figure 5**: The answers in Class 2 on the question "What is the most interesting aspects(s) of playing the game". 13 students answered the questionnaire. The students were allowed to choose more than one of the four options.

These results are interesting because educational games are often viewed as a tool to motivate students to participate in educational activities. Here games become a tool to frame or facilitate processes where the motivation lies in the subject that the game covers or in the research context outside the school context. One student described that the most interesting aspect of playing the game was knowing, "that you have a real chance to help enable a quantum computer. It also irritated me when I didn't get as many points in the game as my friends did, which got me to play the games even more." In response to the same question, two other students explained, "The best thing about the game is that it is one of the few games in this world that you can actually use for something real" and "that the games were relevant to physics, and that you have the chance to make a difference, even if it's not a vital one, in the development of quantum computers (Cool!)". The first answer indicates that the focus that some students have on scores does not necessarily exclude a focus on research collaboration. The two subsequent answers show that it is primarily, if not exclusively, the fact that they are contributing to research that motivates them to play the game.

During the interview conducted immediately after the test in Class 2, 2 - 3 students stated that they did not feel as if they had learned physics. They explained that this was due to the fact that moving the atoms was a simple task that did not expand their understanding of quantum physics. The questionnaire also showed that a larger part of the students did not experience being taught physics through the game to a significant degree. When asked whether they had learned about physics from playing the game using a five-point scale ranging from 1 for "not at all" to 5 for "a lot", most of the students answered 2, and none answered 5, see figure 6.

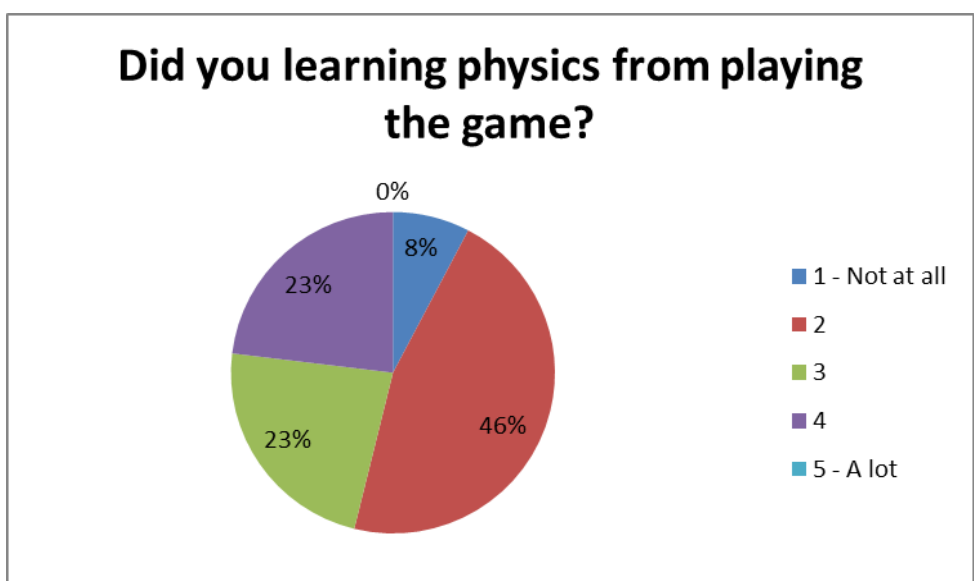

**Figure 6**: The answer in Class 2 on the question "Did you learn physics from playing the game?" on a range from 1 (not at all) to 5 (a lot). 13 students answered the questionnaire.

Although the results seem discouraging, they were no surprise to the development team, since in this early intervention little had been done to align the game simulations with the particular curriculum of the students and the intervention was too short to accommodate extensive learning. Furthermore, in the interview with the teacher after the test session, the teacher challenged the understanding that students had of "learning physics". He argued that their understanding of "learning physics" was to advance their skills in solving concrete problems related to their final exam. He added that in his opinion the contact with the researchers and the game had given his students a deeper understanding of quantum mechanics that he could not have given them as a teacher alone. Although not explicitly tested at the final exam, such "nature of science" is also a part of the curriculum according to the teacher, but understandably the students had a tendency to value learning that prepared them for the final upcoming exam more highly. The conclusion was to aim future intervention design at extending the "nature of science"-like understanding of the concepts of quantum physics, while also incorporating more explicit learning.

In summary, the results from these two classes showed that the main motivating factor was research collaboration or solving physics problems. 54% of students in Class 2 experienced learning no or little physics from playing the game. The game and setup in the class thus provided a strong experience of participating in an authentic experiment, but less evident experience of learning physics from the participation. Teachers in both classes also commented on the "tangible" approach the games had to a highly theoretical subject which is very different from "normal" teaching on this subject. In Denmark, high school physics is taught at a highly theoretical level, which may be the basis for the student's comment that they "only calculate the results" in "normal" teaching, but that the game gave them the feeling of "directly doing the experiment".

**4.3 Test in Class 3: Experimenting with strengthening the student experience of learning**
Class 3 comprised a third-year high school class consisting of 20 students 17-20 years of age from a technical high school in west Denmark.
The feedback and findings in the tests conducted in Class 1 and 2 suggested a different focus for the redesign of EQM. Part of the new focus for the further development of the game was to strengthen the students' experience of participating in an authentic science experiment as this had proven to be a strong motivating factor for some students. Moreover, our hypothesis was that the authentic research collaboration aspect of the game could contribute with new didactical input to science education. One of the ways we strengthened the authentic aspects of the game was to make the researchers more visible on the game's website [scienceathome.org](scienceathome.org). Their pictures were included and the graphics were adjusted to match the atmosphere of the physics lab where data from the game were actually being used to develop a quantum computer.
Another issue that the project team responded to with respect to the second test was making improvements related to the teacher's comment about a tangible approach to a highly theoretically subject. The teacher at Class 3 was interviewed on this and other topics before the test. In this interview she explained that she had observed that this "more intuitive" approach to a highly theoretical subject appealed to a group of students that was exceedingly interested in the subject, but felt that it was difficult. These teacher's statements were in accord with students' comments from Class 1. As a result, the test focused on investigating this issue further.
The final area of focus for implementing and testing the game in Class 3 was to strengthen the students' perceived learning on the subject at both an experimental and theoretical level.  To strengthen the EQM learning environment a collaborative learning design inspired by profession simulation games (Magnussen, 2008) was, as described earlier, implemented in Class 3. Students were divided into three different teams by the teacher dependent on their individual skills. Each team represented an area of expertise: experimental physicists working on understanding movements of atoms better by doing analogous experiments, IT professionals specialised in simulations doing virtual experiments in the game, and theoretical experts focusing on understanding the mathematics behind the various elements of the movement of the atoms. Students worked in groups with similar expertise for three class periods and were then mixed with students with different expertises to share results and to produce a poster or film with their conclusions. Overall the students worked intensively with the game in the different teams. The simulation of the different professional approaches in the authentic framework of contributing to the scientific domain of these professions appeared to spur complex discussions concerning the results obtained from using physical experimental tools compared to the virtual experiments in the game. In the questionnaire students were after the game asked whether they had learned about physics, see the results in figure 6. In Class 3

29 % of the students answer 5 ("a lot"), which is a huge improvement to the result in Class 2 where no one answered 5.

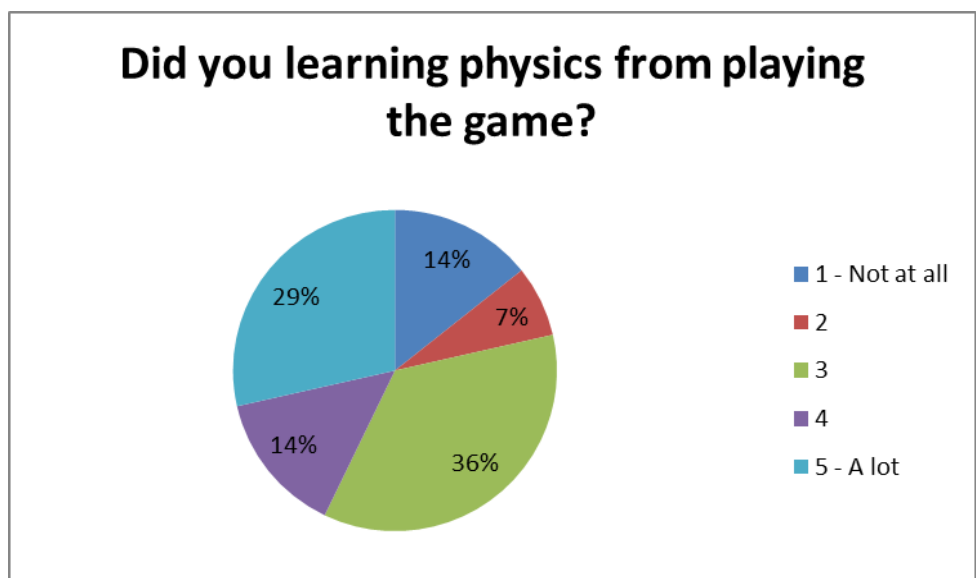

**Figure 6:** The answer in Class 3 on the question "Did you learn physics from playing the game?" on a range from 1 (not at all) to 5 (a lot). 14 students answered the questionnaire

In summary results in Class 3 showed that 79% of students answered 3 or above compared to only 46 % in Class 2. The above results indicate that the new design has strengthened the students' experience of learning physics. This seems to generally for the class, but could potentially be specifically for the less theoretically skilled students. For this group of students EQM offers experimental approaches to quantum mechanics through playing by making direct manipulation atoms possible in the game.

**5. Discussion and Conclusions**
The results presented in this paper show that the educational adaptation of Scientific Discovery Games, enabling participation in authentic scientific experiments, are highly motivating for students. However, the findings also show that the learning design in the class setting must be considered seriously in order to improve the students' experience of learning and that various design challenges remain to be addressed even further. In order to successfully develop and introduce Scientific Discovery Games into science education we need to focus on how the game operates and the didactical aspects that can strengthen important elements in these games, such as authenticity and authentic experimentation. This paper described how a Scientific Discovery Game can be didactically designed to fit a classroom setting by merging aspects from simulation science games about specific professions with the research collaboration approach.
Furthermore, the complexity of playing the game needs to correspond to the complexity of the scientific challenge. This, however, is not straightforward, if even possible, due to the research nature of the game as the "rules" of the research area determines the game mechanics. Another aspect that needs further investigation is how this type of game can motivate weaker students in science education. In interviews with teachers from Class 1 and 3 the teachers stated that the intuitive or tangible approach of the game to quantum physics encouraged weaker students to participate more actively.

The students' responses described in this paper indicate that Scientific Discovery Games offer new approaches to integrate authentic knowledge creation and scientific practice into school science education for specific groups of students. Finally, further research is needed concerning how role playing and collaborative learning approaches can further strengthen students' learning experiences and the outcome of that learning.

**References**

Brown, A. (1992). Design experiments: Theoretical and methodological challenges in creating complex interventions in classroom settings, *The Journal of Learning Sciences*, Vol 2, no. 2, pp141-178.

Cobb, P., Confre, J., diSessa, A., Lehrer, R., & Schauble, L. (2003) Design experiments in education research. *The Educational Researcher*, vol. 32, no. 1, pp. 9-13.

Cooper, S., F. Khatib, I. Makedon, H. Lu, J. Barbero, D. Baker, J. Fogarty, Z. Popović and Foldit Players (2011). Analysis of social gameplay macros in the Foldit cookbook, *Proceedings of the Sixth international Conference on the Foundations of Digital Games,* FDG 2011, Bordeaux, France, pp. 9-14.

Cooper, S., Treuille, A., Barbero, J., Leaver-Fay, A., Tuite, K., Khatib, F., Snyder, A. C., Beenen, M., Salesin, D., Baker, D., Popović, Z. and Foldit players (2010). The challenge of designing scientific discovery games, *Proceedings of the Fifth international Conference on the Foundations of Digital Games*, FDG 2010, New York, NY, pp. 40–47.

Dillenbourg, P. (1999). What do you mean by collaborative learning? in: Dillenbourg, P. (Ed) Collaborative-learning: Cognitive and Computational Approaches, p. 1-19 Oxford: Elsevier.

Egenfeldt-Nielsen, S. (2005). Beyond Edutainment: Exploring the Educational Potential of Computer Games. Copenhagen: IT University of Copenhagen.

Ellington, H., F. Addinall & F. Percival (1981) Games and Simulations in Science Education. London: Kogan Page Ltd.

Grover, L.K. (1996) A fast quantum mechanical algorithm for database search, *Proceedings of the 28th Annual Symposium on the Theory of Computing*, NY, NY: ACM Press, 212-218.

Gee, J.P. (2003) *What video games have to teach us about learning and literacy*. New York: Palgrave Macmillan.

Gee, J. (2005) What would a state of the art instructional video game look like? *Innovate*, vol. 1, no. 6.

Good, B. M. & Su, A. I. (2011) Games with a scientific purpose. *Genome Biology*, vol. 12, no. 135, pp. 1-3.

Kvale, S. (1996) *Interviews—An introduction to qualitative research interviewing*, Thousand Oaks, CA: Sage.

Magnussen, R. (2007) Games as a platform for situated science practice. In de Castell, S., & Jenson, J. (ed.) *Worlds in Play: International Perspectives on Digital Games Research* (301–311). NY, NY: Peter Lang.

Magnussen, R. (2008) *Representational inquiry in science learning games*. Doctoral dissertation, Copenhagen: Aarhus University.

Sawyer, R.K. (2006) Educating for innovation, *Thinking Skills and Creativity*, vol. 1, no. 1, pp. 41-48.

Shaffer, D. W. (2006) Epistemic frames for epistemic games. *Computers & Education*, vol 46 no. 3, pp.223-234.

Shaffer, D. W. (2007). *How Computer Games Help Children Learn*. New York: Palgrave

Sherson, J. F., Weitenberg, C., Endres, M., Cheneau, M., Bloch, I., & Kuhr, S. (2010) Single-atom-resolved fluorescence imaging of an atomic Mott insulator, *Nature*, vol 467, pp. 68–72.

Squire, K.D. (2005) Resuscitating research in educational technology: Using game-based learning research as a lens for looking at design-based research. *Educational Technology* vol. 45, no. 1, pp. 8-14.

Squire, K., & Klopfer, E. (2007) Augmented reality simulations on handheld computers. *Journal of the Learning Sciences*, vol. 16, no. 3, pp. 371-413.

Weitenberg, C., Kuhr, S., Molmer, K. & Sherson, J. F. (2011) Quantum computation architecture using optical tweezers, *Physical Review A*, Vol 84, No. 3, pp. 032322.